\documentclass[prb,preprint,amssymb,amsfonts,superscriptaddress]{revtex4}
\usepackage{epsfig}
\usepackage{amsmath}

\begin{document}

\title{Optical Gain from InAs Nanocrystal Quantum Dots in a
Polymer Matrix}

\author{Gang Chen} 
 \affiliation{Bell Laboratories, Lucent Technologies,
600 Mountain Avenue, Murray Hill, New Jersey 07974}

\author{Ronen Rapaport}
\affiliation{Bell Laboratories, Lucent Technologies, 600 Mountain
Avenue, Murray Hill, New Jersey 07974}

\author{Dan T. Fuchs}
\affiliation{Bell Laboratories, Lucent Technologies, 600 Mountain
Avenue, Murray Hill, New Jersey 07974}

\author{Sahar Vilan}
\affiliation{Bell Laboratories, Lucent Technologies, 600 Mountain
Avenue, Murray Hill, New Jersey 07974}

\author{Assaf Aharoni}
\affiliation{Department of Physical Chemistry and the Farkas
Center for Light Induced Processes, The Hebrew University,
Jerusalem 91904, Israel}
\author{Uri Banin}
\affiliation{Department of Physical Chemistry and the Farkas
Center for Light Induced Processes, The Hebrew University,
Jerusalem 91904, Israel}

\begin{abstract}
We report on the first observation of optical gain from InAs
nanocrystal quantum dots emitting at 1.55 microns based on a
three-beam, time resolved pump-probe technique. The nanocrystals
were embedded  into a transparent polymer matrix platform suitable
for the fabrication of integrated photonic devices.
\end{abstract}

\maketitle

The recent development of chemically
synthesized nanocrystal quantum dots (NQDs) with widely tunable
emission color\cite{Murray2001} has greatly broadened the scope of
the future application of quantum dots in lasers and amplifiers.
Optical gain and amplified stimulated emission from NQDs has
already been observed both in the visible using CdSe
\cite{KlimovScience2000, EislerApplPhysLett2002} and near-IR using
PbSe\cite{Schaller2003} as the core material. A different system
studied for near-IR (around 1.5 $\mu m$ telecommunication
wavelength) applications uses InAs
\cite{Guzelian1996,CaoJAmChemSoc2000}, which, compared to PbSe,
has a much lower degree of degeneracy for the lowest quantized
states \cite{KangJOSAB1997}. It could therefore be a promising
candidate for reducing gain threshold.

It is also important to find a convenient way to integrate NQDs
into future planar optical circuits. We have recently developed a
low-loss polymer platform in which the InAs as well as PbSe NQDs
were dispersed into perfluorocyclobutane (PFCB) polymer with their
optical properties preserved \cite{Olsson2004}. Optical devices
can then be made after thermally curing spin-casted NQD-polymer
films\cite{Olsson2004}. Here we present the first experimental
evidence for the observation of optical gain in an
InAs-NQD-polymer film ready for device fabrication at room
temperature. Using a two- and a three-beam time-resolved
pump-probe techniques, we measure the gain dynamics of the InAs
NQDs, extracting their gain lifetime and the gain recovery time.

\begin{figure}[t!]
\begin{center}
\includegraphics[scale=0.8]{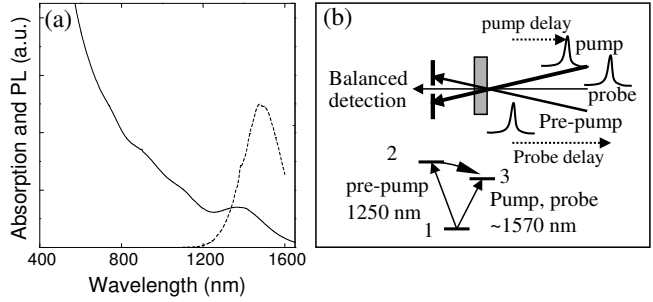}
\caption{(a) Absorption (solid line) and photoluminescence (dashed
line) spectra of the InAs NQD-polymer film. (b) A schematic
diagram of the three-beam pump-probe setup and the level diagram
for the NQDs. For the conventional two-beam pump-probe
measurements, the resonant pump beam is blocked.}\label{figure1}
\end{center}
\end{figure}

PFCB \cite{SmithAdvMat2002} films containing InAs NQDs were
fabricated following the procedure described in Ref.
\cite{Olsson2004}. The NQD weight fraction in the polymer host was
experimentally measured to be 1\% (based on optical technique
mentioned in Ref. \cite{Olsson2004}), corresponding to an InAs
volume fraction of 0.27\%. For an average NQD size of 8 nm
(estimated from TEM images), it corresponds to an NQD density of
$1.0\times 10^{16}cm^{-3}$. The Film thickness was measured to be
0.6mm. Typical absorption and photoluminescence (PL) spectra at
room temperature are shown in Fig.~\ref{figure1}(a) and are not
affected by the NQD-polymer mixing process\cite{Olsson2004}.

\vspace*{0cm}

The optical measurements were performed at room temperature and
were based on a three-beam pump-probe technique\cite{KimAPL2002}
schematically shown in Fig.~\ref{figure1} (b). The optical pulses
($\sim$120fs) were generated using two independently wavelength
tunable optical parametric amplifiers in the near IR pumped by a
single 800 nm Ti:Sapphire regenerative amplifier. The optical
pulses generated from the first OPA were used as a pre-pump. The
pulses from the second OPA were spit into a pump and a probe beam.
The time delays between each of the three light pulses were
controlled using two motorized delay stages. The three beams were
focused on the film using long focal length lenses such that their
overlap length is comparable to the sample thickness. In a typical
measurement, the probe intensity was maintained extremely low.


This technique offers a convenient way to measure small internal
optical gain even in the presence of appreciable device losses.
Here, the pre-pump is tuned to 1250 nm and excites the NQDs from
the ground state (level 1 of Fig.~\ref{figure1} (b)) to higher
excited states (level 2). The excited carriers then quickly relax
nonradiatively to the lowest confined levels of NQDs (level 3) to
possibly give rise to a population inversion. The pump and probe
are tuned to be resonant with the transition between the level 1
and level 3 at 1570 nm and pass through the sample shortly after
the excitation by the pre-pump. The differential transmission of
the probe due to the resonant pump is measured. In the case that
no population inversion between level 3 and 1 is induced by the
pre-pump, the probe experiences saturated absorption without the
resonant pump. Because the resonant pump excites even more
carriers to level 3, the probe experience even less absorption due
to the saturation of the 3-1 transition. The differential probe
transmission signal in this case would be positive following the
resonant pump. When there is population inversion, however, the
probe experiences a gain without the resonant pump. The presence
of the resonant pump would only reduce the population inversion
via stimulated emission, leading to less probe gain. The
differential probe transmission signal in this case is therefore
negative immediately after the resonant pump. Thus, by studying
the sign of such a differential signal, one can unambiguously
determine whether the NQDs are population inverted, even in the
presence of strong losses.

\begin{figure}[t!]
\begin{center}
\includegraphics[scale=1.0]{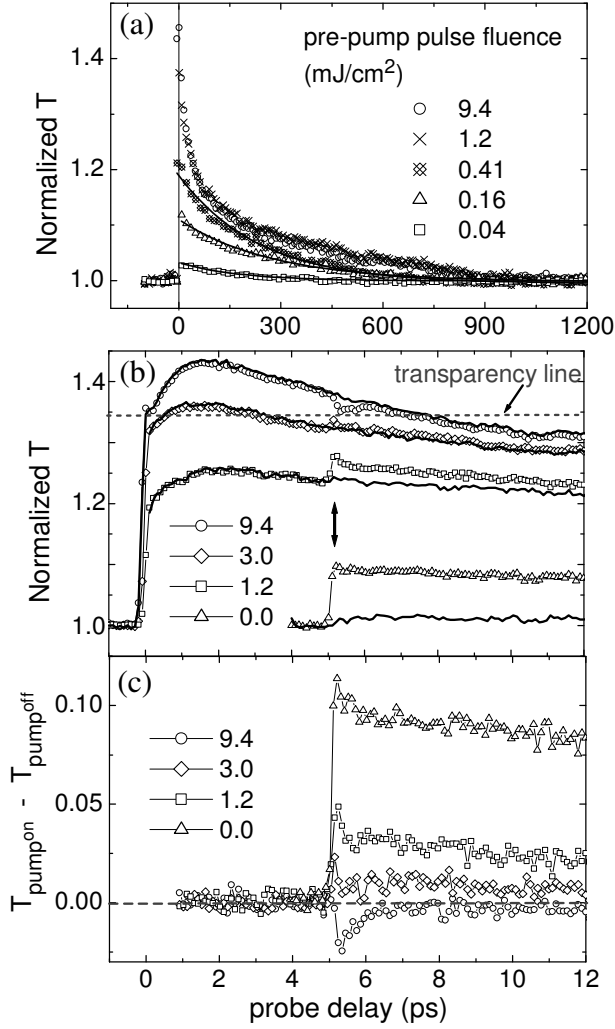}
\caption{(a) Normalized probe transmission as a function of the
pre-pump - probe delay for different pre-pump fluences in the
two-beam geometry in which the resonant pump is not present. (b)
Same as (a), but now the resonant pump pulse is used and is
delayed by 5ps with respect to the pre-pump (marked by the arrow).
The dark lines present the corresponding two-beam pump-probe
experimental traces with the resonant pump off. The dashed line
(transparency line) marks the onset of gain. (c) Same data as in
(b) presented as differential probe transmission due to the
resonant pump.} \label{figure2}
\end{center}
\end{figure}

Fig.~\ref{figure2} (a) presents the probe transmission traces as a
function of the delay between the probe and the pre-pump for
various pre-pump fluences (in units of mJ/cm$^2$, fluence is for
each pulse) in the common two-beam pump-probe experimental
geometry (with the resonant pump blocked). The signals are
normalized to their values when the pre-pump fluence is zero.
These traces show that after the excitation by the pre-pump, level
3 is quickly populated via nonradiative relaxation of carriers
from level 2 to 3, and the probe experiences an increase in
transmission, as expected. As the pre-pump intensity is turned up,
the level 3 population following the pre-pump increases and so
does the probe transmission signal. At low pre-pump excitation
intensity, the probe transmission after the excitation of the
pre-pump follows a single exponential decay with a decay constant
of $\sim 300ps$ (carrier radiative lifetime), shown by the curve
fitting for traces with pre-pump fluences equal to or smaller than
0.4 mJ/cm$^2$. Ideally, this corresponds to an excitation of less
than an average of 1 e-h pair per NQD \cite{KlimovScience2000}.
Indeed, the number of excited e-h pairs for 0.4 mJ/cm$^2$ pre-pump
is estimated to be 0.8, using an NQD density of $1.0\times
10^{16}cm^{-3}$ (see above), a measured beam size of 100 $\mu m$
and a measured pre-pump absorption of 30\%. For higher pre-pump
excitation fluences that excites more than 1 e-h pairs per NQD,
the probe transmission decay clearly becomes multi-exponential
with the faster components determined by the enhanced nonradiative
Auger relaxation \cite{KlimovScience2000-2}.

\vspace*{0cm}

Fig.~\ref{figure2} (b) presents the probe transmission traces as a
function of the delay between the probe and the pre-pump and as a
function of the pre-pump fluence for the three-beam measurements.
The resonant pump is delayed by 5ps (marked by the arrow) and is
maintained a constant intensity (fluence of 1.2 mJ/cm$^2$). The
probe transmission is normalized to its value when all pumps are
off. The dark lines represent traces where the resonant pump is
off, similar to the two-beam pump-probe experiments discussed
previously. In these measurements, whether the population between
level 1 and 3 is inverted is determined by the change of the probe
transmission immediately after the excitation by the resonant
pump. At low pre-pump intensity, level 3 is less populated than
level 1, the resonant pump thus drives level 3 occupancy higher,
leading to increased transmission of the probe. This shows up as
an additional peak on top of the probe transmission traces at 5 ps
where the resonant pump arrives. At high pre-pump intensity,
however, level 3 and 1 population is inverted. The resonant pump
reduces the inversion via stimulated emission. The probe thus
experiences less gain and its transmission dips at the arrival of
the resonant pump. The dotted line in Fig.~\ref{figure2} (b) marks
the probe transmission level where the dip and peak tend to meet.
It represents the transparency level of the probe.

The differential probe transmission due to the resonant pump (the
symboled traces in Fig.~\ref{figure2} (b) minus the corresponding
dark line traces) is plotted in Fig.~\ref{figure2} (c), showing
the sign change between the gain and no-gain cases. The sign
change occurs roughly at a pre-pump fluence of 3 mJ/cm$^2$, with
15\% absorbed by the NQD film (measured), corresponding to an
excitation of 6 e-h pairs per NQD (see above for parameters used).
Since the gain threshold occurs at 1 e-h pair at the lowest
confined levels per NQD, this suggests that the quantum efficiency
of exciting electrons into the lowest confined states by the
pre-pump is about 20\%. The maximum internal gain achieved in
these measurements was 2/cm, which corresponds to an NQD gain
cross section of $2\times 10^{-16}cm^2$ for a NQD density of
$1.0\times 10^{16}cm^{-3}$. We note that much higher gain is
expected by decompressing the pre-pump pulse to increase
excitation efficiency, by improving the NQD quantum efficiency,
and by increasing the NQD loading in the polymer host.

\vspace*{0cm}
\begin{figure}[t!]
\begin{center}
\includegraphics[scale=0.8]{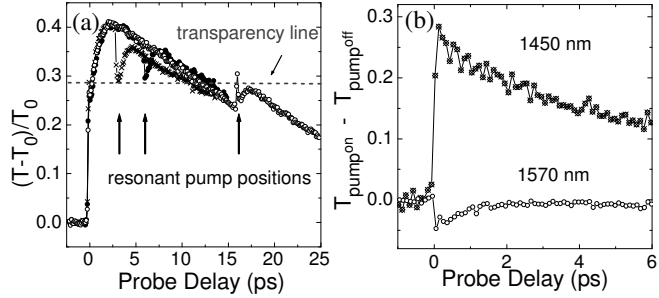}
\caption{(a) Normalized probe transmission with different resonant
pump delays, marked by the arrows. The dashed line (transparency
line) marks the onset of gain. The pre-pump fluence is 9.4
mJ/$cm^2$. (b) Probe Differential transmission for two different
probe and resonant pump wavelengths with a pre-pump fluence of 9.4
mJ/$cm^2$ and wavelength of 1250nm.} \label{figure3}
\end{center}
\end{figure}

The measurements above show that immediately after the excitation
by a strong pre-pump, level 3-1 population is inverted. The
carrier decay in the NQDs, however, leads to a reduction and
eventually loss of the population inversion after the pre-pump
passes the sample. By delaying the resonant pump and probe pulses
different amount of time relative to the pre-pump, the gain
lifetime can be measured. This is shown by the probe transmission
signal in Fig.~\ref{figure3} (a) for a pre-pump and pump fluence
of 9.4 and 3.0 mJ/cm$^2$ respectively. The arrows indicate the
position of the resonant pump, For short resonant pump delay
times, a negative differential signal of the probe transmission is
observed, indicating that the probe experiences an internal gain.
At longer delays, the differential signal is positive, which
indicates that the population inversion and gain is lost. The
probe transmission tends to approach the transparency line (doted
line) immediately after the resonant pump for all pre-pump-pump
delays. From the crossing between the decay curves and the
transparency line, the gain lifetime can be inferred to be
$\sim$10ps. The short gain lifetime corresponds to the fast
initial carrier decay when the NQDs are occupied by more than 1
e-h pairs. The dominant source for this decay is likely to be the
nonradiative Auger recombination \cite{KlimovScience2000-2}.

The carrier dynamics after the excitation by the resonant pump can
be extracted from Fig.~\ref{figure2} (c). For low pre-pump
intensities (no-gain regime), the positive differential probe
transmission signal decay similar to that of the high power
two-beam pump-probe signal, with the decay rate determined by the
Auger as well as the radiative carrier recombination.  For high
pre-pump intensities (gain regime), the negative differential
probe transmission signal recovers on a time scale of $\sim$2 ps,
much faster than the carrier recombination time. In this case, the
resonant pump first reduces the level 3 population via stimulated
emission, effectively depleting the probe gain instantaneously.
The probe gain then recovers. This gain recovery can be explained
by other highly excited electrons quickly replenishing the NQDs
depleted by the resonant pump to recover the level 3 occupancy.
Gain recovery was also observed in other systems and its time
scale is a very important parameter lasers and amplifiers.

Finally, we show in Fig.~\ref{figure3} (b) traces of the
differential probe transmission as a function of the
probe-pre-pump delay under the same high pre-pump fluence (9.4
mJ/cm$^2$) for two different wavelengths of the probe and the
resonant pump. Again, the resonant pump is delayed by 5 ps with
respect to the pre-pump. While at 1570nm the probe experiences
gain, it does not experience any net gain and shows a positive
differential signal at 1450 nm. This confirms that population
first occurs at the low energy side of the inhomogeneously
broadened spectrum of the NQD ensemble.

In summary, we have successfully incorporated InAs NQDs emitting
at 1.55 $\mu m$ into a PFCB polymer platform. We show for the
first time optically induced population inversion and gain from an
PFCB film containing InAs NQDs based on a three-beam pump-probe
technique. The gain lifetime and recovery time were studied. The
authors thank Xing Wei and Phil Platzman for helpful discussions.

\end{document}